\documentclass{JFM-FLM_Au}
\usepackage{graphicx}
\graphicspath{{.//}}
\usepackage{amsmath, amssymb}
\usepackage{natbib}
\usepackage{color}
\usepackage{hyperref}

\usepackage{enumerate} 
\usepackage{amssymb} 

\usepackage{comment}
\usepackage[nameinlink]{cleveref}
\crefname{figure}{Figure}{Figures}
\Crefname{figure}{Figure}{Figures}
\crefname{equation}{Eq.}{Eqs.}
\Crefname{equation}{Eq.}{Eqs.}
\crefname{section}{Sec.}{Sections}
\Crefname{section}{Section}{Sections}
\usepackage{subcaption}

\lefttitle{P. M\"uller, K. Ihrig, A. Sack, J.E. Kollmer, and T. P\"oschel}
\righttitle{Journal of Fluid Mechanics}

\title{Rapid generation of relaxed centimeter-sized drops by capillary auto-ejection}
\author{
  Patric M\"uller\aff{1}, 
  Kai Ihrig\aff{1}, 
  Achim Sack\aff{1}, 
  Jonathan E. Kollmer\aff{2} \and 
  Thorsten P\"oschel\aff{1}
}
\date{\today}

\affiliation{
\aff{1}Institute for Multiscale Simulation, Friedrich-Alexander Universität Erlangen-Nürnberg, 91058 Erlangen, Germany
\aff{2}German Aerospace Center (DLR), Institute for Frontier Materials on Earth and in Space, Department Functional Granular and Composite Materials, Cologne, Germany
}

\corresau{Thorsten P\"oschel, \email{thorsten.poeschel@fau.de}}

\usepackage[dvipsnames]{xcolor}
\usepackage{etoolbox}
\usepackage{placeins}

\begin{document}
\maketitle

\begin{abstract}
Many microgravity experiments require centimeter-sized spherical liquid drops. In drop-tower experiments, only a few seconds of microgravity are available, so the drops must be generated and reach a nearly relaxed state within a very short time. We investigate capillary auto-ejection (puddle jumping) as a passive mechanism for producing free-floating drops and investigate how the supporting cavity can be shaped to minimize the relaxation time after ejection. 

For cavities with a parabolic cross-section filled with a fixed liquid volume, minimization of the initial liquid surface area yields a unique shape, and smoothed particle hydrodynamics simulations show that this shape also gives the shortest relaxation time within the parabolic family. Introducing a shallow central depression into the cavity reduces the relaxation time further, despite producing a larger initial liquid surface area. These results provide design principles for capillary-driven drop generators that enable the rapid production of relaxed centimeter-sized drops during the short microgravity intervals available in drop towers. 
\end{abstract}


\section{Introduction}
Many experiments in fluid dynamics, interfacial science, combustion, and materials science rely on macroscopic, nearly spherical liquid drops as reproducible model systems. Such drops serve as controllable structures for exploring capillary phenomena, interfacial instabilities, oscillations, evaporation, coalescence, and wetting phenomena. They also act as canonical test cases for validating numerical models of capillary and interfacial flow, with direct applications in coating, propellant management, and space-borne manufacturing \citep{Dietrich2022, Frank2019, Smirnov2020, Wollmann2023}. 

Under terrestrial conditions, gravity inevitably deforms large drops. Microgravity environments offer a possibility to overcome this limitation. While platforms such as the International Space Station enable extensive experimentation, access to these facilities is limited. In contrast, drop towers provide high-quality microgravity conditions with the same residual gravity as the ISS ($10^{-6}\,g$), however, only for a few seconds \citep{Kobel2020, Wollmann2023}. Consequently, the available time for drop generation, relaxation, and subsequent measurements is severely restricted. Any reduction in the time required to produce a stable spherical drop directly increases the useful experimental time. Experiments performed under such conditions therefore require drop-generation mechanisms that operate rapidly and reproducibly.

\begin{figure}[t!]
    \centering
    \includegraphics[width=0.99\textwidth]{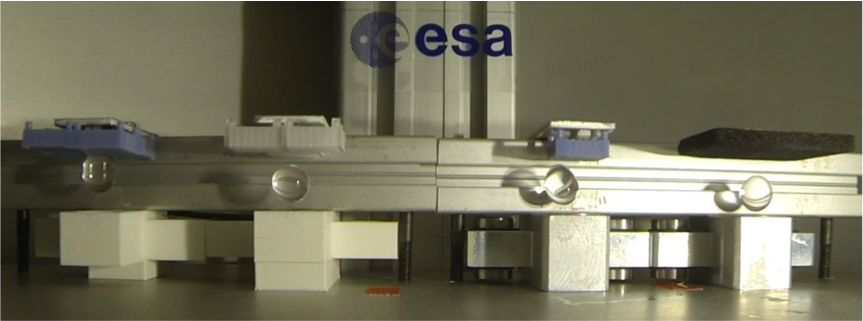}
    \caption{Snapshot of four drops immediately after auto-ejection from the hemispherical cup, showing near-spherical shape during free flight, in a drop-tower experiment.}
    \label{fig:snapshot}
\end{figure}

Capillary auto-ejection, commonly referred to as \emph{puddle jumping}, provides a passive mechanism for generating freely suspended drops, see \cref{fig:snapshot}. In this process, a liquid confined within a hydrophobic cavity spontaneously detaches once curvature-induced capillary pressure exceeds the confining forces \citep{Wollman2012}. 
Following ejection, the drop relaxes toward its spherical equilibrium shape under the action of surface tension. For short-duration microgravity experiments, the usefulness of this mechanism is therefore determined not only by reliable drop ejection but also by how rapidly the drop relaxes.

The relaxation process is governed by the initial shape of the liquid immediately before ejection, which in turn is determined by the shape of the supporting cavity. This raises the question of how the cavity should be designed to minimize the relaxation time. A natural hypothesis is that the optimal cavity is the one that minimizes the initial liquid surface area at fixed volume, since surface tension tends to drive the drop toward a sphere. Whether this geometric criterion is sufficient, however, remains unclear because the subsequent relaxation is a fully dynamic process involving inertia, viscosity, and capillary forces.

In this work, we investigate the relationship between cavity shape and drop relaxation following capillary auto-ejection. We first consider cavities with a parabolic cross-section, for which the geometry minimizing the initial liquid surface area can be obtained analytically. Smoothed particle hydrodynamics simulations are then used to determine the corresponding relaxation times and to assess whether geometric optimization indeed minimizes the relaxation time. Finally, we extend the study to more general cavity shapes and demonstrate that introducing a shallow central depression leads to substantially faster relaxation than is achievable with parabolic cavities. The resulting design principles enable the rapid generation of relaxed centimeter-sized drops within the short microgravity intervals available in drop-tower experiments.

\section{Optimization of Trough Shape}
\subsection{Roadmap and motivation}
Surface tension drives the liquid toward configurations of smaller surface area, ultimately favouring a spherical equilibrium shape. This suggests that cavity shapes producing an initial liquid configuration with a small surface area may reduce the relaxation time after ejection.

We therefore begin by considering parabolic troughs, for which the shape minimizing the initial liquid surface area can be determined analytically. Whether this geometric optimum also minimizes the relaxation time, however, cannot be inferred from geometry alone because the subsequent evolution is governed by the full fluid dynamics, and dynamic effects must be resolved explicitly. We therefore employ smoothed particle hydrodynamics simulations to determine the relaxation time, verify the geometric optimum, and extend the optimization to more general cavity shapes.

\subsection{Geometric optimization of a parabolic trough\label{sec:parabolic}}
\begin{figure}[t!]
    \centering
    \includegraphics[width=0.4\textwidth]{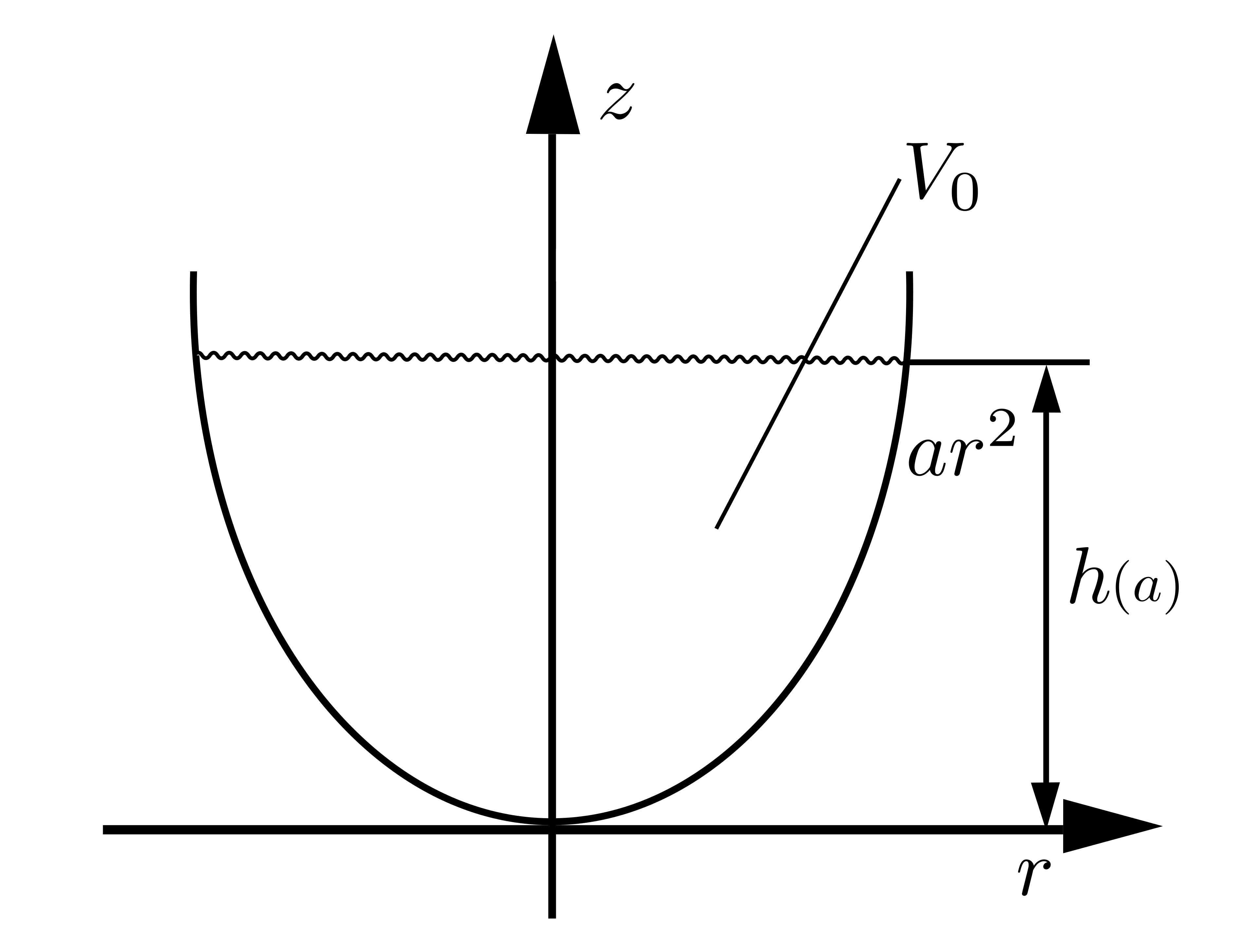}
    \caption{Sketch of the parabolic trough shape. }
    \label{fig:sketch-parabolic}
\end{figure} 

We first consider the one-parameter family of axisymmetric parabolic troughs, which constitutes the simplest one-parameter family of smooth cavity shapes. The trough profile is described by
\begin{equation}
    z = a r^2,\qquad a>0,
\end{equation}
where $a$ denotes the curvature parameter. Rotating this profile about the $z$-axis yields an axisymmetric paraboloidal cavity. The cavity is filled with a prescribed liquid volume $V_0$ up to the height $h(a)$ (\cref{fig:sketch-parabolic}). The total liquid surface consists of a planar circular disk of radius $R(a)=\sqrt{h/a}$ at $z=h$, and the curved paraboloid surface extending from $z=0$ to $z=h$:
\begin{equation}
S_\text{tot}(a)=
\pi R(a)^2 + 2\pi\int_0^{R(a)} r\sqrt{1+4a^2r^2}\,dr ,
\end{equation}

The enclosed volume up to height $h$ is
\begin{equation}
V(h,a)=\int_0^{h(a)} \pi r(z)^2\,dz=\frac{\pi}{2a}{h(a)}^2\,.
\end{equation}
Since the optimization is carried out at fixed volume the problem contains only one independent dimensionless combination of the curvature parameter $a$ and the liquid volume $V_0$. We therefore introduce
\begin{equation}
 k \equiv 4\sqrt{\frac{2V_0}{\pi}}\,a^{3/2}\,.
\end{equation}
With this definition, the total surface area can be written in the scale-invariant form
\begin{equation}
S_\text{tot}(a)=V_0^{2/3}\,\Phi(k)\,,
\end{equation} 
where the dimensionless function is given by
\begin{equation}
\Phi(k) = \frac{\pi^{1/3}2^{1/3}}{3}\;
\frac{3k + 2(1+k)^{3/2} - 2}{k^{4/3}}.
\end{equation}

Minimizing $S_\text{tot}(a)$ at fixed $V_0$ is equivalent to minimizing the dimensionless $\Phi(k)$. Analytical minimization yields the unique positive root
\begin{equation}
k_\text{opt} = 4\left(3 + \sqrt{3}\right).
\label{eq:k_opt}
\end{equation}
Inverting the definition of $k$ gives the optimal curvature parameter in dimensional form,
\begin{equation}
a_\text{opt}
= \left(\frac{k_\text{opt}}{4\sqrt{2/\pi}}\right)^{2/3}\; V_0^{-1/3}
\equiv C\,V_0^{-1/3},
\end{equation}

with the universal numerical constant
\begin{equation}
{\,C= \sqrt[3]{3\pi \left(2 + \sqrt{3}\right)}\approx 3.276\,.}
\end{equation}

\subsection{Dynamic relaxation: Smoothed Particle Hydrodynamics Simulations}
The geometric optimization identifies the paraboloidal cavity with the smallest initial liquid surface area. Whether this geometry also minimizes the relaxation time, however, must be established from the fluid dynamics. We therefore perform SPH simulations of the complete ejection and relaxation process.

The relaxation following capillary auto-ejection involves large deformations of the liquid interface, rapid changes in curvature, and moving contact with the cavity during the early stages of ejection. Numerical simulations, therefore, require a method capable of accurately resolving strongly deforming free surfaces together with capillary forces. Traditional Eulerian methods require explicit interface reconstruction and are less suited to such strongly deforming interfaces. We therefore employ smoothed particle hydrodynamics (SPH), a fully Lagrangian mesh-free method in which the liquid interface is represented naturally by the outermost particles. This makes SPH particularly well suited for simulating capillary auto-ejection and the subsequent drop relaxation. Only the essential elements of the SPH formulation are summarized below; a detailed description of the numerical model is given in \citep{Blank_2024}.

\subsubsection{Particle approximation of the Navier-Stokes-Equations} 
For an incompressible fluid, the governing equations are
\begin{align}
\rho\frac{D\vec{u}}{Dt} &= -\nabla p + \mu \nabla^2\vec{u} + \vec{f}^\mathrm{s} + \rho\vec{g}\,, 
\label{eq:NS}\\
\frac{\partial \rho}{\partial t} + \nabla\cdot(\rho\vec{u}) &= 0 \quad\Rightarrow\quad \nabla\cdot\vec{u}=0\quad \text{(incompressible)}\,, 
\label{eq:continuity}
\end{align}
where $\vec{u}$ is the velocity, $\rho$ the density, $p$ the pressure, $\mu$ the dynamic viscosity, $\vec{f}^\mathrm{s}$ denotes surface-tension forces and $\vec{g}$ is the gravity.

The material derivative
\begin{equation}
    \frac{D\vec{u}}{Dt}=\frac{\partial\vec{u}}{\partial t}+(\vec{u}\cdot\nabla)\vec{u}
\end{equation}
is the acceleration of a material (infinitesimal) fluid element. SPH adopts a Lagrangian viewpoint: the continuum is replaced by a finite set of material elements (SPH particles) that carry mass $m_a$, position $\vec{r}_a$, velocity $\vec{u}_a$ and other field quantities. Each particle therefore follows Newton's equation,
\begin{equation}
m_a\frac{d\vec{u}_a}{dt} = \vec{F}_a\,.    
\end{equation}

\subsubsection{Kernel approximation of continuum fields}
The evaluation of the forces $\vec{F}_a$ requires the continuum fields appearing in \cref{eq:NS}—-pressure, velocity gradients, and viscous stresses—-to be known at the particle locations.

In SPH, these continuum fields are reconstructed from the discrete particle data using kernel-based integral interpolation \citep{gingold1977sph,lucy1977sph,monaghan2005smoothed}. Any continuum field $\phi(\vec{r})$ is approximated by 
\begin{equation}
    \phi(\vec{r}) \;\approx\; \int_{\Omega} \phi(\vec{r}')\, W(\vec{r}-\vec{r}',h)\, d\vec{r}'
    \;\;\longrightarrow\;\;
    \sum_b \frac{m_b}{\rho_b}\, \phi_b\, W_{ab},
    \label{eq:sph-interp}
\end{equation}
where $W_{ab}=W(\vec{r}_a-\vec{r}_b,h)$ is a compactly supported smoothing kernel of width $h$, and the sum runs over all particles within the kernel support. Spatial derivatives are obtained by differentiating the kernel,
\begin{equation}
    \nabla \phi(\vec{r}_a)
    \;\approx\;
    \sum_b \frac{m_b}{\rho_b}\, \phi_b\, \nabla_a W_{ab}\,.
    \label{eq:sph-grad}
\end{equation}
This allows all differential operators in \cref{eq:NS} to be expressed directly in terms of particle quantities. We employ the Wendland $C^2$ kernel \citep{wendland1995piecewise,dehnen2012sph}, defined by
\begin{equation}
W(r,h) \;=\;
\begin{cases}
\displaystyle
\frac{21}{16\pi h^3}\left(1 - \frac{r}{2h}\right)^{4}\left(1 + 2\frac{r}{h}\right), 
& 0 \le r < 2h, \\[10pt]
0, 
& r \ge 2h ,
\end{cases}
\end{equation}
The Wendland kernels are widely used in modern SPH due to their improved stability, compact support, and excellent pairing resistance \citep{dehnen2012sph}, providing smooth and robust field reconstructions.

Using this kernel interpolation, the pressure contribution on the right-hand side of the Navier-Stokes equation \cref{eq:NS} is discretized following the symmetric formulation by \citet{monaghan2005smoothed}, which ensures momentum conservation and suppresses tensile instabilities:
\begin{equation}
\vec{F}^\text{(p)}_a=-\nabla p
= - \sum_{b} m_b 
\left(
\frac{p_a}{\rho_a^2} + \frac{p_b}{\rho_b^2}
\right)\,\nabla_a W_{ab}.
\label{eq:pressure-force}
\end{equation}

Viscous forces are modelled using the SPH discretization by \citet{Viscosity_Morris_1997}, which is appropriate for Newtonian fluids:
\begin{equation}
\vec{f}^\mathrm{v}_{a} = \sum_b \frac{m_b\left(\eta_a + \eta_b\right) \vec{u}_{ab}}{\rho_a\rho_b} F_{ab},\quad \vec{u}_{ab} \equiv \vec{u}_a - \vec{u}_b\,,
\end{equation}
where $\vec{u}_a$ and $\vec{u}_b$ are the velocities of particle $a$ and $b$ and $\eta_a$ and $\eta_b$ are the dynamic viscosity coefficients assigned to particles $a$ and $b$. The factor $F_{ab}$ is given by
\begin{equation} F_{ab} \equiv F\left(\vec{r}_{a}-\vec{r}_{b}\right) = \frac{\vec{r}_{ab} \cdot \vec{\nabla} W_{ab} }{r_{ab}^2 + \left(0.01h\right)^2}, \quad \vec{r}_{ab}\equiv \vec{r}_a-\vec{r}_b. \end{equation}
The small regularization term $\left(0.01h\right)^2$ prevents divergence when two particles approach each other.

\subsubsection{Surface tension and free surface treatment} 
Near free surfaces, the standard SPH density formulation underestimates $\rho_a$ because neighboring particles are missing, leading to spurious pressure gradients. Particles near the free surface are identified using the Shepard filter \citep{ShepardFilter}
\begin{equation}
S_a = \sum_b \frac{m_b}{\rho_b} W_{ab}.
\end{equation}
When $S_a$ falls below a prescribed threshold, kernel support is deficient. 
In this case, the pressure forces are corrected according to \citet{Blank_2024}. This ensures that particles at the liquid–gas interface satisfy the correct pressure boundary condition without explicitly simulating the gas phase:
\begin{equation} 
\vec{f}^\mathrm{p}_{a} =  -\sum\limits_{b} m_{b} \left( \dfrac{p_b- p^{0}_{a} }{\rho_{b}^2} \right) \vec{\nabla} W_{a{b}}
\end{equation}
where $p_a^0$ denotes the prescribed external pressure corresponding to the Dirichlet boundary condition at particle $a$. 
Surface tension is modeled using the continuum surface force (CSF) formulation \citep{brackbill1992continuum}, which requires the interface normal and curvature.  The interface normal is computed from the kernel gradient,
\begin{equation}
\vec{n}_a = -\sum_b \frac{m_b}{\rho_b}\nabla W_{ab}.
\end{equation}
The smoothed, normalized interface normal
\begin{equation} \tilde{\vec{n}}_a = \frac{1}{S_a}\sum_{b} \frac{m_b}{\rho_b} \vec{n}_b W_{ab}
\end{equation}
is then used to evaluate the curvature $\kappa_a$:
\begin{equation}
\kappa_a =-\frac{1}{2} \sum_b\frac{m_b}{\rho_b} \left(\hat{\tilde{\vec{n}}}_a - \hat{\tilde{\vec{n}}}_b \right) \cdot \vec{\nabla} W_{ab}.
\end{equation}
For constant surface tension $\sigma$, the resulting surface-tension acceleration is
\begin{equation}
\vec{f}^\mathrm{s}_a
= \frac{2\sigma}{\rho_a}\,\kappa_a\,\tilde{\vec{n}}_a\,\delta^\mathrm{s}_a.
\end{equation}
The surface delta function $\delta^\mathrm{s}_a$ distributes the interfacial surface-tension force over a thin volumetric layer surrounding the liquid-gas interface \citep{Blank_2024}.

\subsubsection{Incompressible smoothed particle hydrodynamics}
To enforce incompressibility $\nabla\cdot\vec{u}=0$, we employ the predictor-corrector scheme proposed by \citet{cummins1999sph}: the position of particle $a$ at time step $n$ is predicted according to
\begin{equation} 
\vec{r}_{a}^\ast= \vec{r}_{a}^n + \vec{u}_{a}^n \Delta t\,;
\quad a=1,2,3,\dots \, . 
\end{equation}
The predicted velocity is obtained by integrating the viscous, surface-tension and body forces (here gravity), over one time step
\begin{equation} 
\vec{u}_{a}^\ast= \vec{u}_{a}^n + \left(\vec{f}_{a^\ast}^\mathrm{v} + \vec{f}_{a^\ast}^\mathrm{s} +\vec{f}_{a^\ast}^\mathrm{b}\right)\Delta t 
\quad \text{with}
\quad \vec{f}_{a^\ast}^\mathrm{v} \equiv \vec{f}^\mathrm{v}\left(\vec{r}^\ast_a\right)\,, 
\quad \text{etc.} 
\end{equation}
The pressure field $p_{a}^\ast$ required to enforce incompressibility is obtained from the pressure Poisson equation (PPE)
\begin{equation} 
\vec{\nabla} \cdot \left( \frac{\vec{\nabla} p_{a}^\ast}{\rho_{a}}\right) 
= \frac{\vec{\nabla} \cdot \vec{u}_{a}^\ast}{\Delta t}\,. 
\end{equation}
Using the predicted velocity $\vec{u}_{a}^\ast$ and the pressure $p_{a}^\ast$, the divergence-free velocity at the next time step $n+1$ reads
\begin{equation} 
\vec{u}_{a}^{n + 1} = \vec{u}_{a}^\ast- \frac{1}{\rho_{a}}\vec{\nabla} p_{a}^\ast\Delta t\,. 
\end{equation}
The corrected particle positions are given by
\begin{equation} 
\vec{r}_{a}^{n + 1} = \vec{r}_{a}^n + \left(\frac{\vec{u}_{a}^{n} + \vec{u}_{a}^{n + 1}}{2}\right) \Delta t\,. 
\end{equation}
Within the SPH framework, the pressure Poisson equation is discretized according to \citet{cummins1999sph} and \citet{monaghan2005smoothed} as
\begin{equation} 
\sum_b \frac{m_b}{\rho_b} \frac{4}{\rho_a + \rho_b} \left(\,p_a^\ast - p_b^\ast\right) F_{ab} 
= \sum_b -\frac{m_b}{\rho_b} \frac{\vec{u}^\ast_{ab} \cdot \vec{\nabla} W_{ab}}{\Delta t}\,,
\label{eq:PPESPH} 
\end{equation}
where $\vec{u}^\ast_{ab} \equiv \vec{u}^\ast_{a} - \vec{u}^\ast_{b}$.
\cref{eq:PPESPH} forms a linear system for the unknown particle pressures $p^\ast_a$, which is solved using the BiCGSTAB iterative solver \citep{bicgstab}.

\subsection{Verification of geometric optimization}
The geometric optimization in \cref{sec:parabolic} predicts an optimal dimensionless curvature parameter of $k_\text{opt} = 18.928$ (see \cref{eq:k_opt}). This prediction is now assessed using SPH simulations. The simulations use the material properties of water at $10^\circ\mathrm{C}$. The liquid volume is fixed at $V_0=1\,\mathrm{ml}$, corresponding to a discretization of 63,335 SPH particles. Because pressure gradients become very small near equilibrium, convergence is assessed from the deviation between the instantaneous liquid surface area and that of a sphere with the same volume. The relaxation time is defined as the first instant at which this deviation falls below 2\%.

\begin{figure}[ht!]
    \centering
    \includegraphics[width=0.8\textwidth]{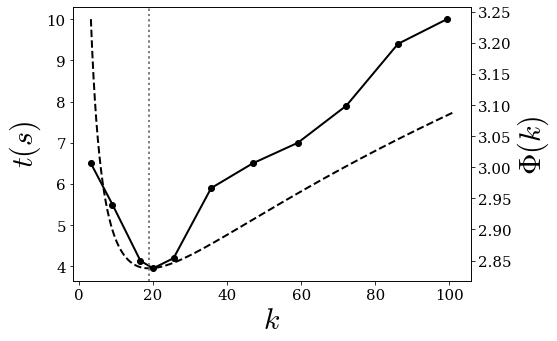}
    \caption{Relaxation time $t$ obtained from SPH simulations (solid line) and dimensionless initial liquid surface area $\phi$ (dashed line) as functions of the dimensionless shape parameter $k$ for parabolic troughs. The vertical dotted line indicates the analytically predicted shape parameter that minimizes the initial liquid surface area.
    }
    \label{fig:relaxation-parabolic}
\end{figure}
\Cref{fig:relaxation-parabolic} compares the analytical prediction with the SPH simulations. The relaxation time exhibits a clear minimum at the analytically predicted value of $k$, confirming that, within the family of parabolic troughs, the cavity minimizing the initial liquid surface area also minimizes the relaxation time. The corresponding minimum relaxation time is approximately $3.9\,\text{s}$.

\begin{figure}[ht!]
    \centering
    \includegraphics[width=0.8\textwidth]{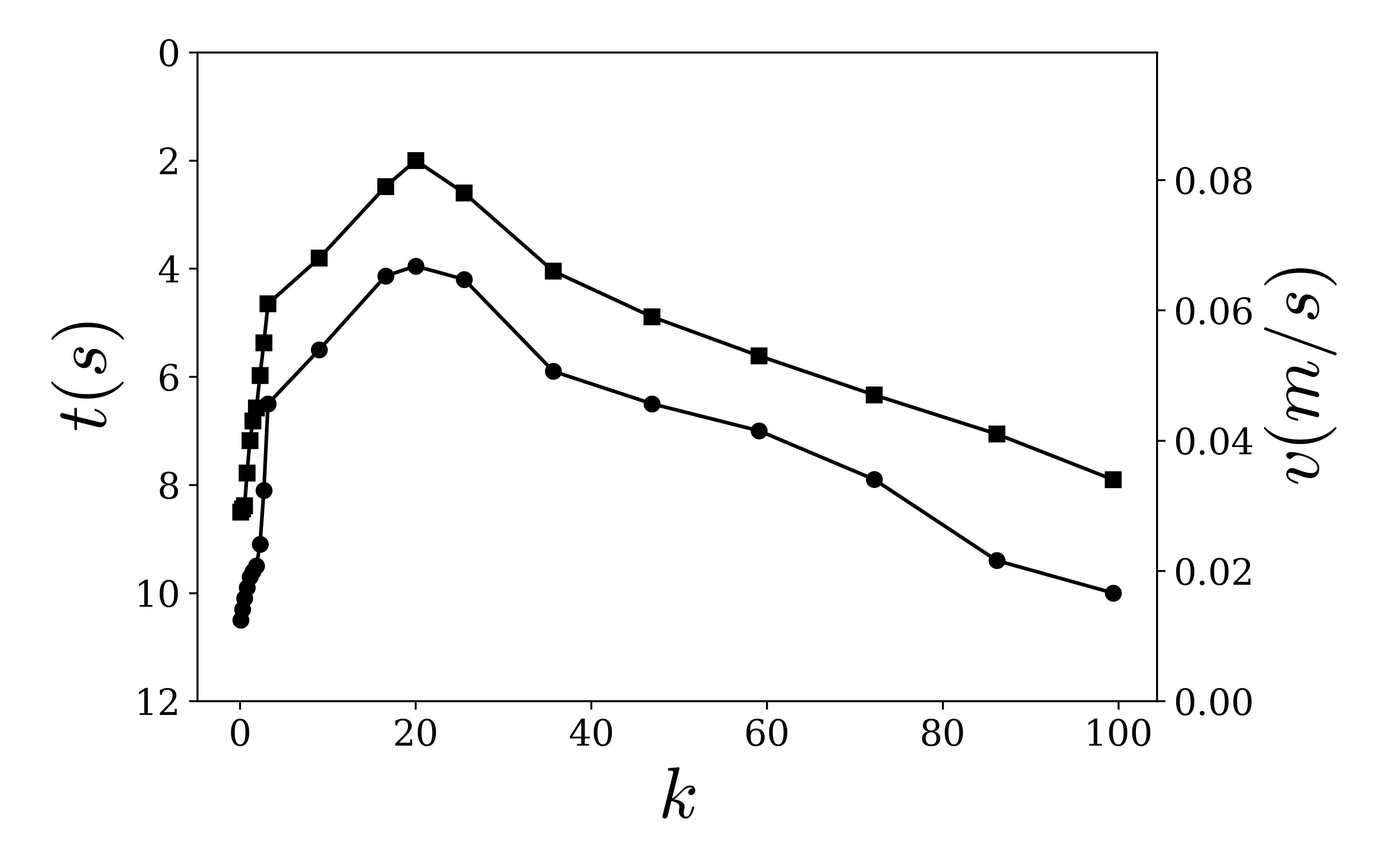}
    \caption{Center-of-mass velocity of the drop immediately after auto-ejection (square symbols) and relaxation time (circle symbols) as functions of the dimensionless shape parameter $k$.}
    \label{fig:com-parabolic}
\end{figure}

\Cref{fig:com-parabolic} shows the center-of-mass velocity immediately after ejection together with the corresponding relaxation time. The shortest relaxation time is obtained for the cavity shape that also produces the largest center-of-mass velocity after auto-ejection, indicating that cavity shapes promoting efficient conversion of surface energy into translational motion also reduce residual shape oscillations.

\subsection{Dynamic optimization of a double-well trough}
\begin{figure}[h]
    \centering
    \begin{subfigure}{0.24\linewidth}
        \raisebox{0.3cm}{\includegraphics[width=\linewidth, trim=43cm 20cm 43cm 0cm, clip]{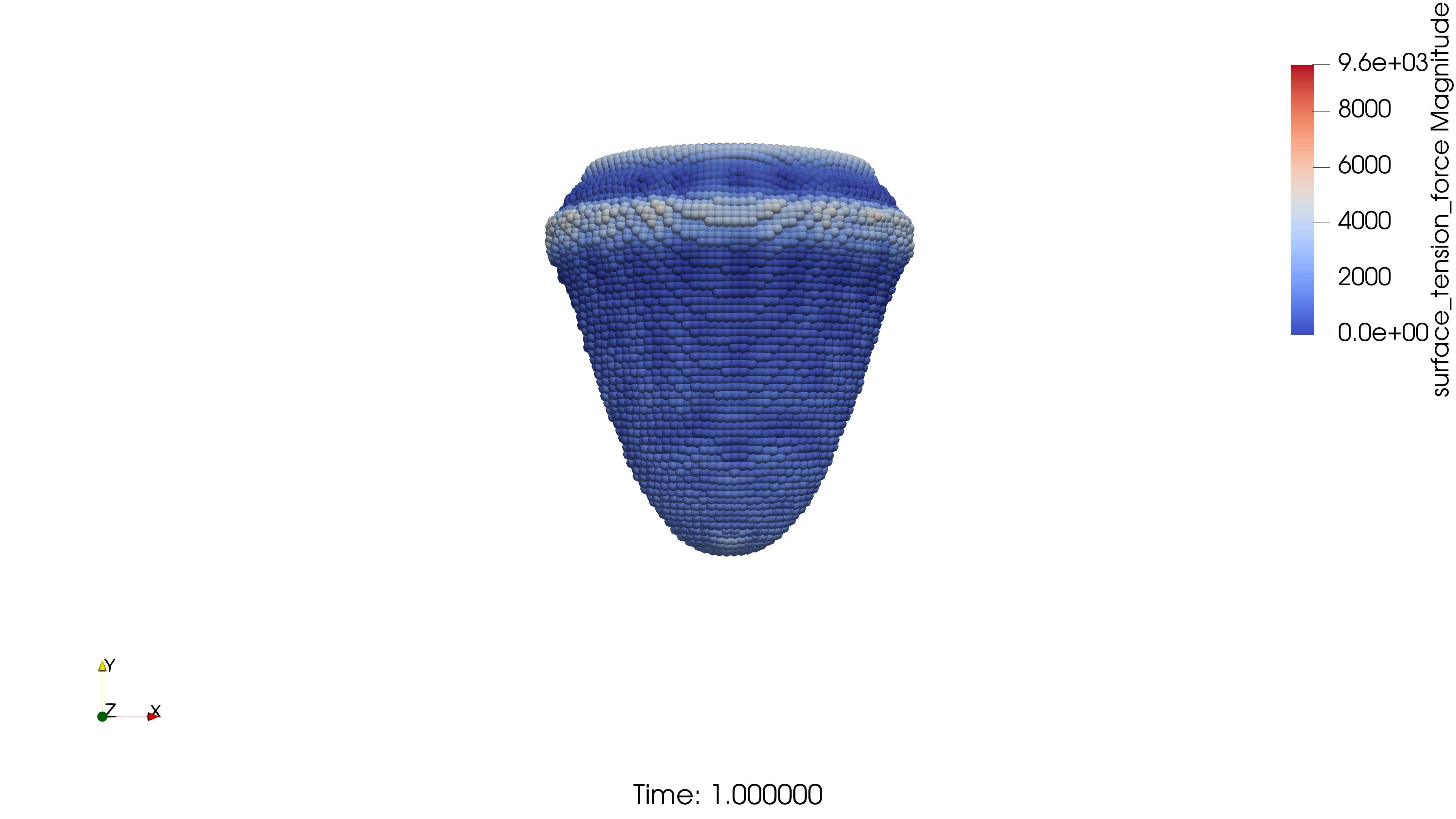}}
    \end{subfigure}
    \begin{subfigure}{0.24\linewidth}
        \includegraphics[width=\linewidth, trim=43cm 20cm 43cm 0cm, clip]{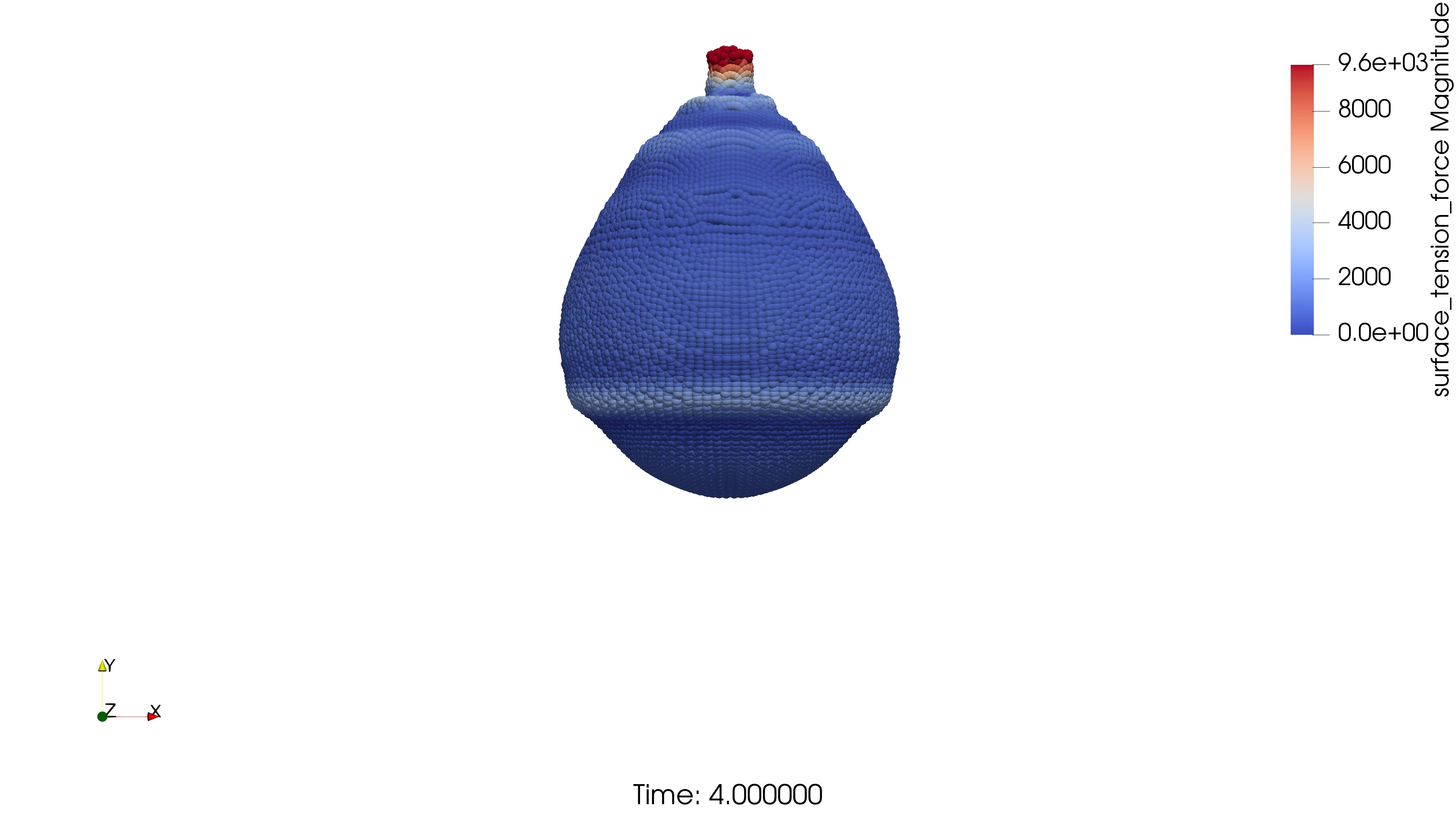}
    \end{subfigure}
    \begin{subfigure}{0.24\linewidth}
        \raisebox{-0.1cm}{\includegraphics[width=\linewidth, trim=43cm 20cm 43cm 5cm, clip]{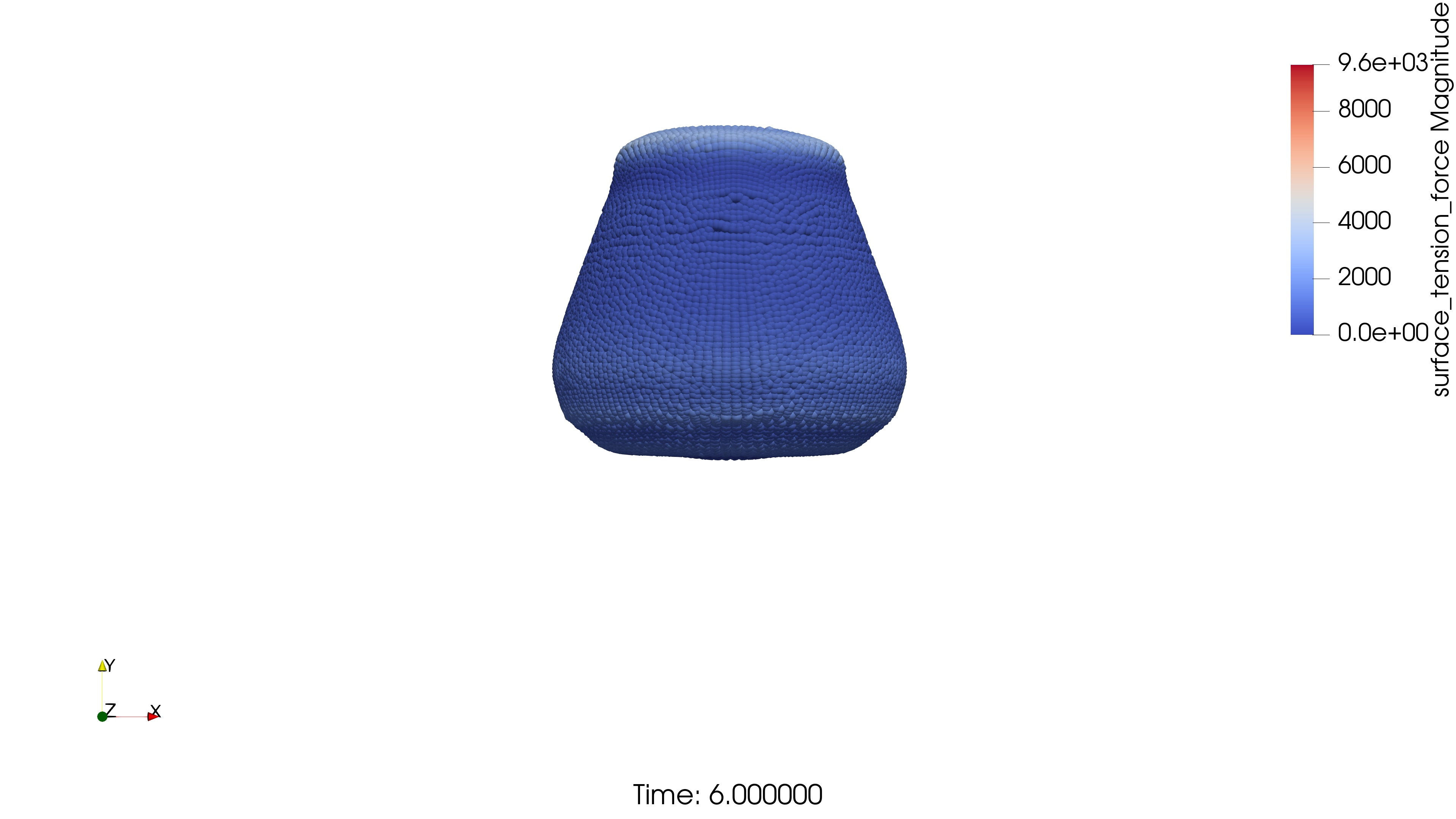}}
    \end{subfigure}
    \begin{subfigure}{0.24\linewidth}
        \includegraphics[width=\linewidth, trim=43cm 20cm 43cm 5cm, clip]{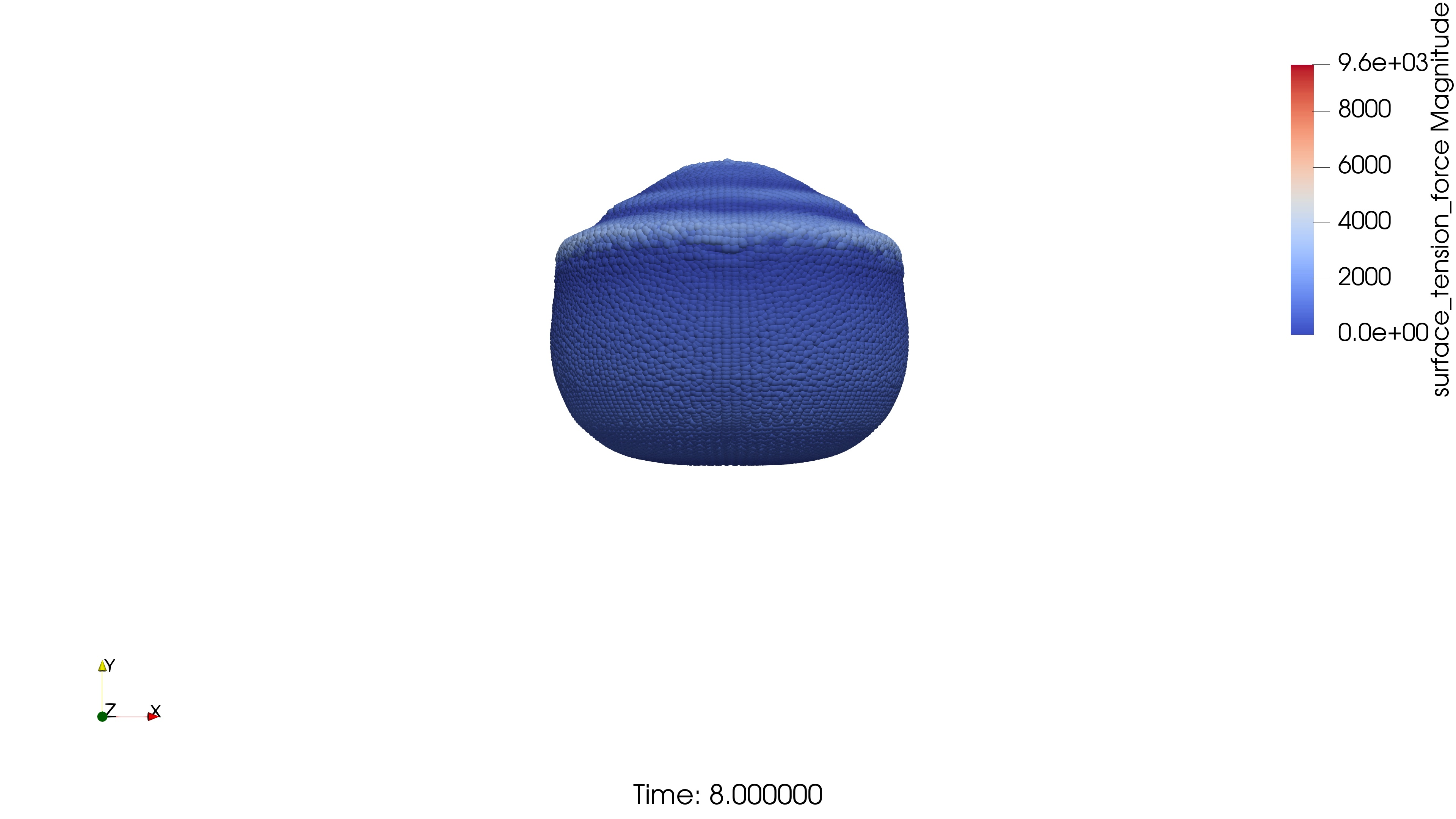}
    \end{subfigure}
    \caption{Sequence of the beginning of relaxation process from a parabolic trough shape 0.005s, 0.02s, 0.03s and 0.04s, with pronounced elongation along the symmetry axis.
    }
    \label{fig:sequence-parabolic}
\end{figure}

\Cref{fig:sequence-parabolic} shows that drops generated from the optimized parabolic trough undergo pronounced elongation along the symmetry axis during the early stages of relaxation. This observation suggests that the initial curvature distribution promotes axial stretching during the early relaxation phase.

To counteract this tendency, we introduce a simple two-parameter family of axisymmetric double-well troughs with a central indentation. The purpose of this shape family is not to identify the globally optimal trough shape, but to investigate how introducing a single additional shape degree of freedom influences the relaxation dynamics. This simple extension of the parabolic trough provides a minimal framework for studying the interplay between cavity shape and transient relaxation.


The objective is therefore no longer to minimize the initial liquid surface area but to suppress the axial deformation that develops during ejection. Consequently, the geometric optimization is replaced by a fully dynamic optimization based on SPH simulations: for each candidate geometry we determine the hydrostatic filling height corresponding to a fixed fluid volume and subsequently perform an SPH ejection simulation. The search is restricted to a simple two-parameter family of axisymmetric trough shapes with a central indentation. This choice introduces the additional shape variation required to suppress axial stretching while keeping the optimization problem computationally tractable.

\begin{figure}[ht]
    \centering
    \includegraphics[width=0.5\textwidth, trim=0cm 0cm 0cm 3.1cm, clip]{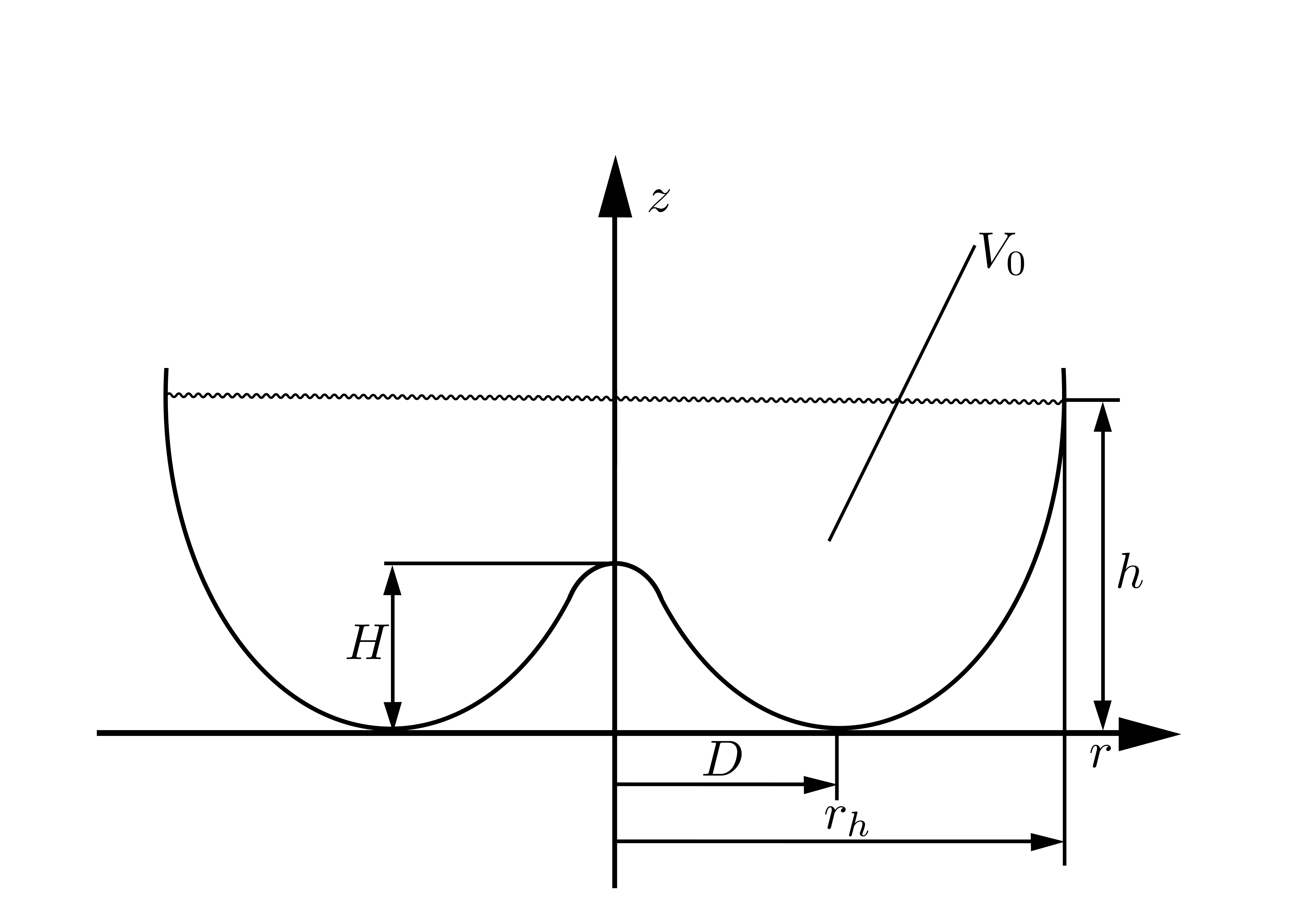}
    \caption{Sketch of the double-well trough.}
    \label{fig:sketch-guglhupf}
\end{figure} 
To keep the parameter space as small as possible while introducing an additional geometric degree of freedom, the profile of the cylindrically symmetric cavity is represented by the cubic polynomial with a central indentation, which is then rotated around the $z$ axis to obtain the three-dimensional cavity shown in \cref{fig:sketch-guglhupf}:
\begin{equation}
  z(r) = H\!\left[
    2\!\left(\frac{r}{D}\right)^3
    - 3\!\left(\frac{r}{D}\right)^2
    + 1
  \right],
  \label{eq:cubic_profile}
\end{equation}
with the geometric constraints being
\begin{align}
z(0) &= H,  & z'(0) &= 0,\\
z(D) &= 0,  & z'(D) &= 0 .
\end{align}
For a given free-surface height, $h$, the radial extent of the liquid is determined by the unique intersection
\begin{equation}
  z(r_\mathrm{h})=h , \qquad r_\mathrm{h} \ge D ,
  \label{eq:sh}
\end{equation}
because $z(r)$ is strictly increasing on $r\ge D$. The enclosed fluid volume is given by
\begin{equation}
  V(h,H,D)
  = 2\pi \int_{0}^{r_\mathrm{h}}
      r \,\bigl(h - z(r)\bigr)\,\mathrm{d}r .
\label{eq:volume_shell}
\end{equation}
To ensure comparability between different geometries we fix the total fluid volume $V$. For each pair $(H,D)$ the fill height $h(H,D)$ is therefore determined implicitly by the constraint
\begin{equation}
  V(h(H,D), H, D)
  = V_{\mathrm{0}} .
  \label{eq:volume_constraint}
\end{equation}
Using the cubic form \cref{eq:cubic_profile}, we change variables to
\begin{align}
    \sigma = \frac{r}{D}, \qquad \hat{\sigma}= \frac{r_\mathrm{h}}{D}
\end{align}

so that the free--surface height satisfies
\begin{equation}
  \frac{h}{H} = 2\hat{\sigma}^3-3\hat{\sigma}^2+1,
\end{equation}
obtained by evaluating \cref{eq:cubic_profile} at $r=r_\mathrm{h}$.

Substituting $r=D\sigma$, $\mathrm{d}r=D\,\mathrm{d}\sigma$, and the cubic expression for $z(r)$ into \cref{eq:volume_shell}, the volume becomes

\begin{align}
  V &= 2\pi\int_{0}^{D\hat{\sigma}} r\bigl(h-z(r)\bigr)\,\mathrm{d}r \nonumber\\
    &= 2\pi D H \left[
      \bigl(2\hat{\sigma}^3-3\hat{\sigma}^2+1\bigr)\frac{\hat{\sigma}^2}{2}
      - \left(\frac{2}{5}\hat{\sigma}^5-\frac{3}{4}\hat{\sigma}^4+\frac{1}{2}\hat{\sigma}^2\right)
    \right] = \pi H D^2 J(\hat{\sigma})\, . 
    \label{eq:V_intermediate}
\end{align}

Since $J(\hat{\sigma})$ is strictly increasing for $\hat{\sigma}\ge 1$, the equation
\begin{equation}
  \pi H D^2 J(\hat{\sigma}) = V_0
\end{equation}

has a unique physically relevant solution $\hat{\sigma}(H,D)$. Once this is found numerically, the fill height follows from
\begin{equation}
  h(H,D) = H\left( 2\hat{\sigma}^3-3\hat{\sigma}^2+1\right)\,,
\end{equation}
and the free-surface radius is $r_\mathrm{h} = D\,\hat{\sigma}$.

\begin{figure}
    \centering
    \begin{subfigure}{0.24\linewidth}
        \includegraphics[width=\linewidth, trim=45cm 10cm 45cm 5cm, clip]{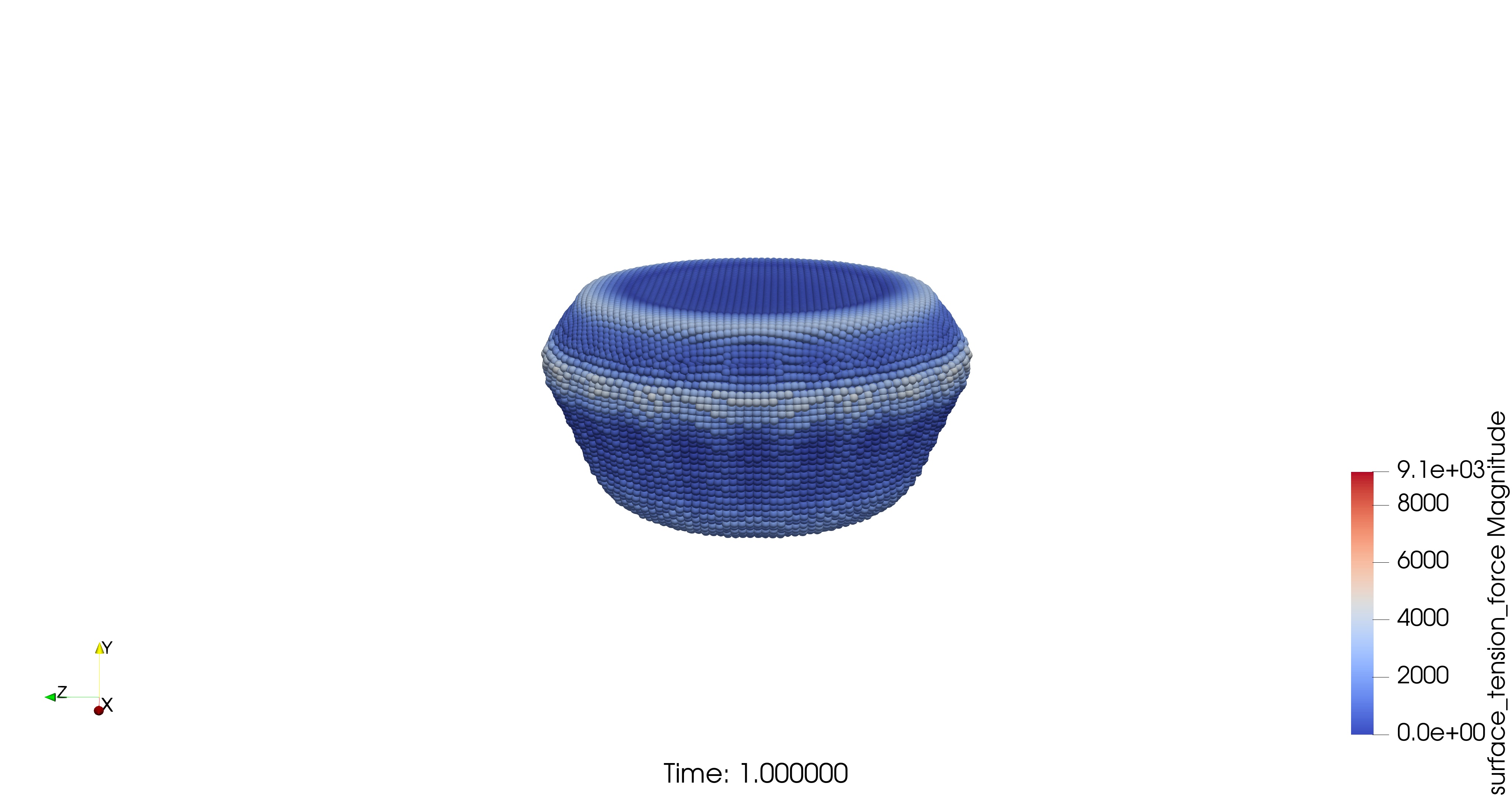}
    \end{subfigure}
    \begin{subfigure}{0.24\linewidth}
        \includegraphics[width=\linewidth, trim=45cm 10cm 45cm 5cm, clip]{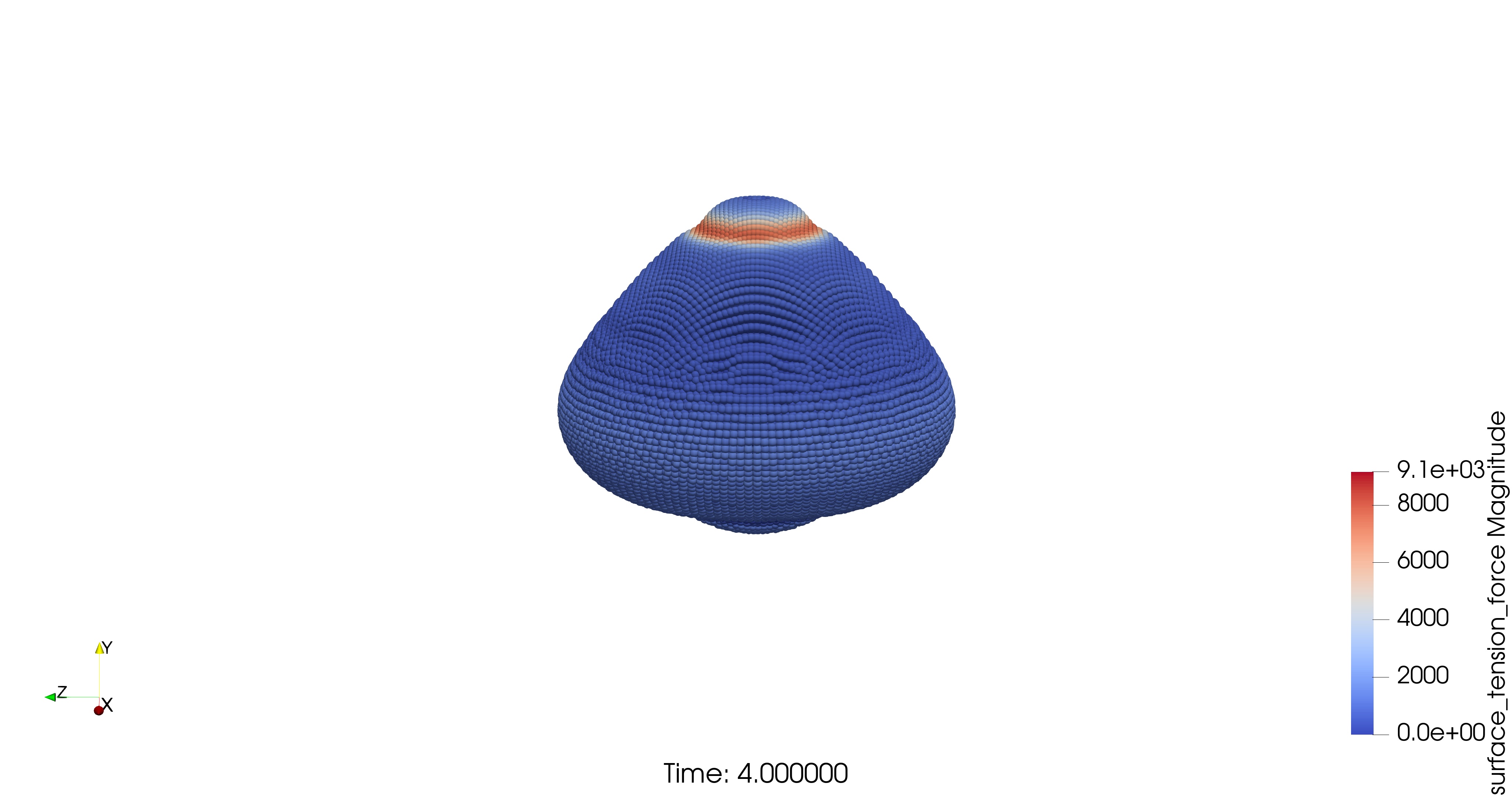}
    \end{subfigure}
    \begin{subfigure}{0.24\linewidth}
        \raisebox{0.1cm}{\includegraphics[width=\linewidth, trim=45cm 10cm 45cm 5cm, clip]{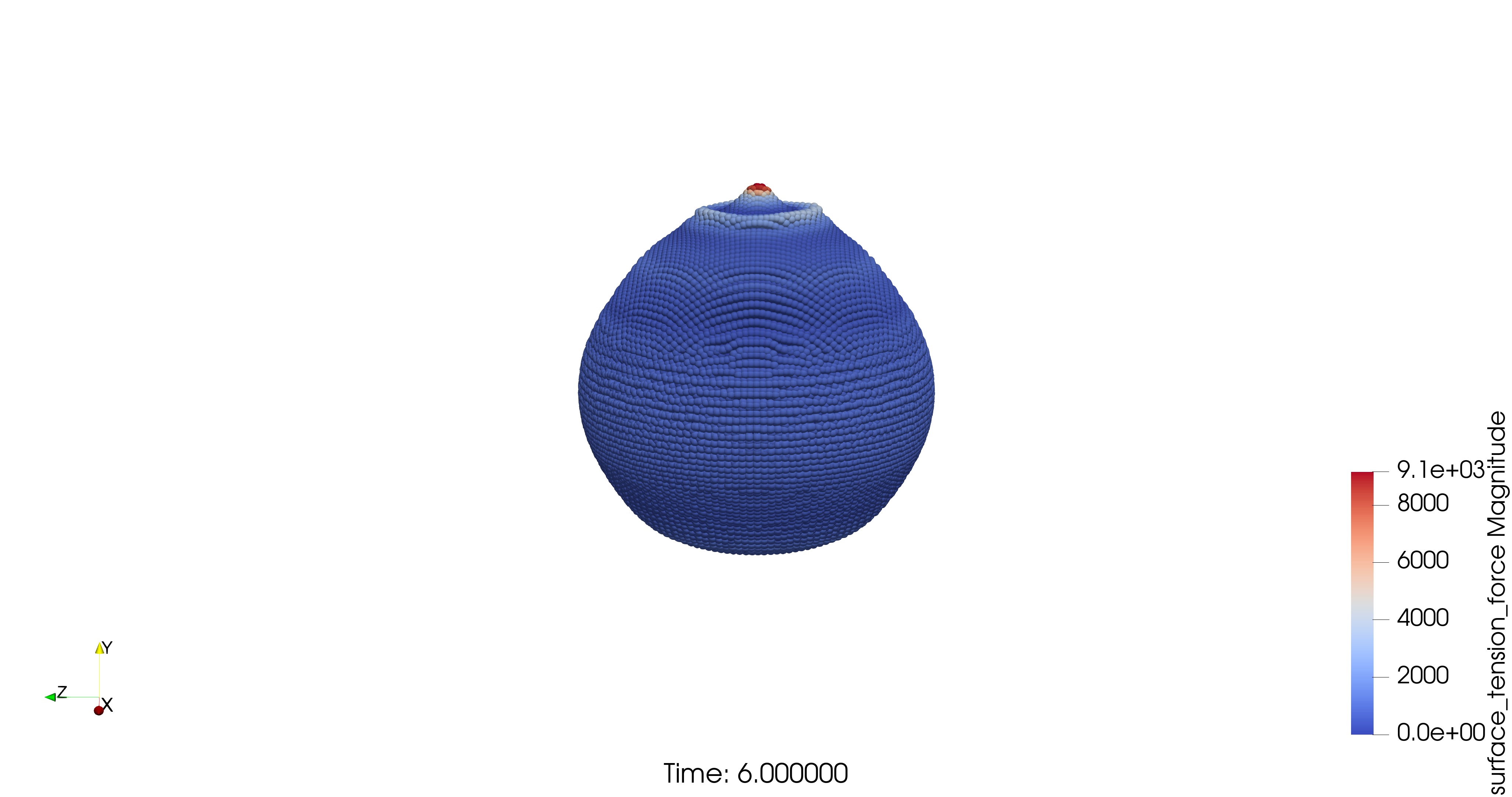}}
    \end{subfigure}
    \begin{subfigure}{0.24\linewidth}
        \raisebox{0.22cm}{\includegraphics[width=\linewidth, trim=45cm 10cm 45cm 5cm, clip]{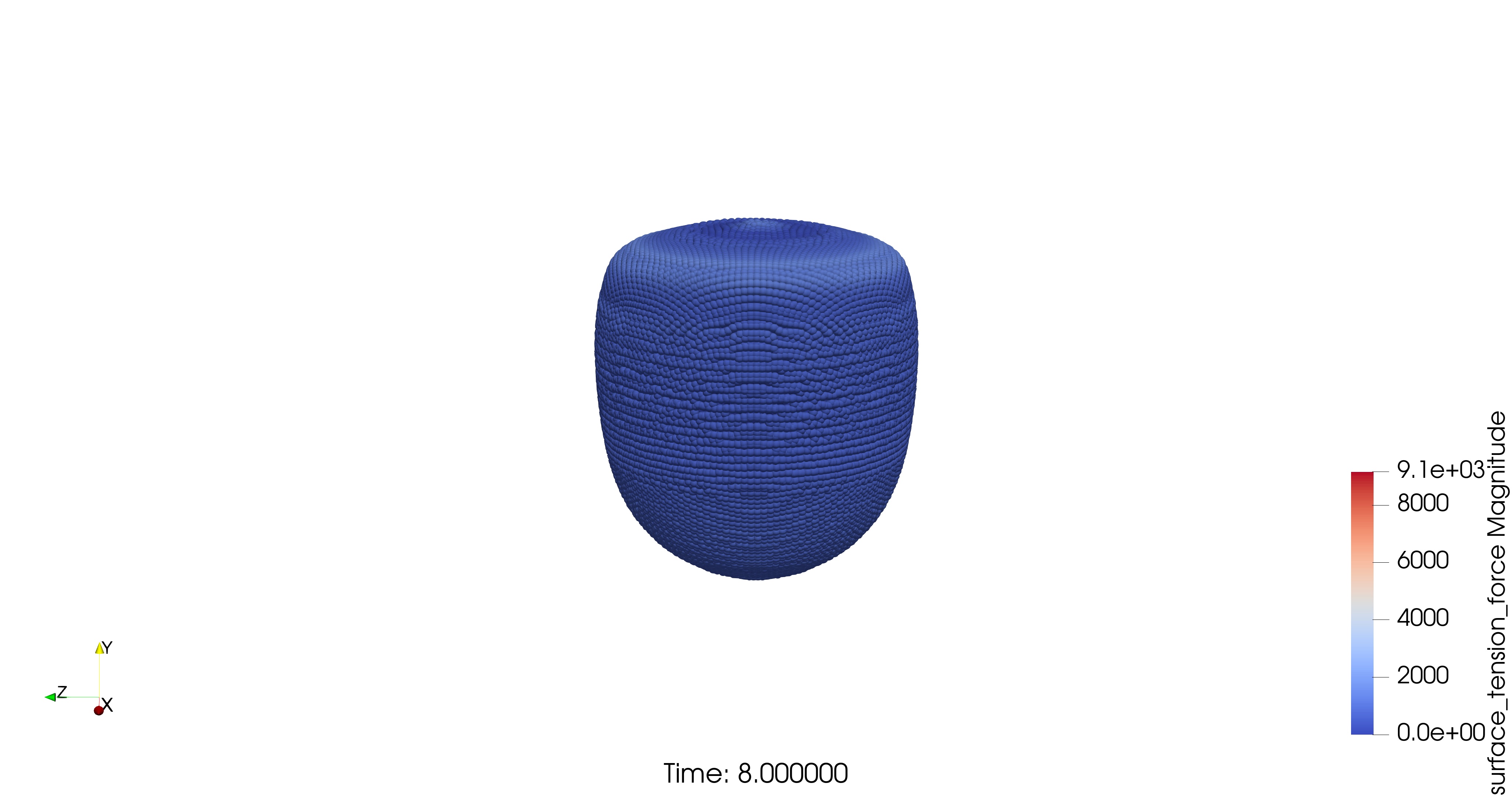}}
    \end{subfigure}
    \caption{Sequence of the beginning of the auto-ejection process at 0.005s, 0.02s, 0.03s and 0.04s from a double-well trough shape with reduced axial elongation.
    }
    \label{fig:sequence-double-well}
\end{figure}

As shown in \cref{fig:sequence-double-well}, the elongation along the symmetry axis can be significantly reduced compared to \cref{fig:sequence-parabolic}, thereby reducing the relaxation time.

\begin{figure}[ht!]
    \centering
    \includegraphics[width=0.4\textwidth]{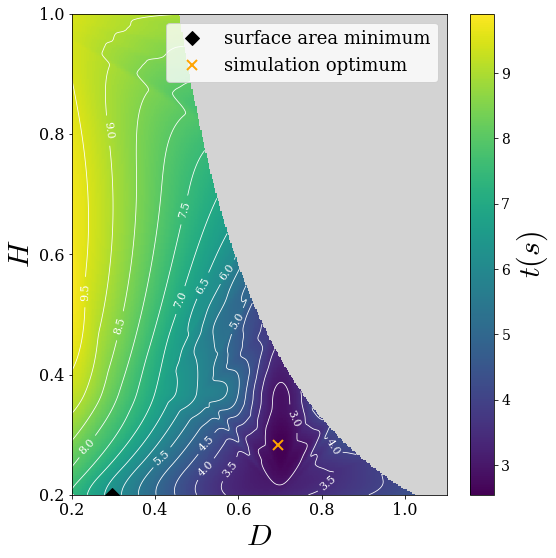}
    \caption{Relaxation time $t$ (color coded) for the double-well trough shape obtained from SPH simulations as a function of the double-well parameters $H$ (height of indentation) and $D$ (width). White contour lines indicate the corresponding equilibrium filling height. The white region is inaccessible because the corresponding parameters yield a filling height below $H$, resulting in an undesired toroidal initial drop shape.
}
    \label{fig:relaxation-double-well}
\end{figure}

Within the parabolic family, minimizing the initial liquid surface area also minimizes the relaxation time. This correspondence no longer holds for the double-well geometry. This shows that, once an additional geometric degree of freedom is introduced, minimizing the initial liquid surface area is no longer sufficient to minimize the relaxation time, and we conclude that dynamic effects dominate over purely geometric surface minimization. This trend is evident from the heat map (\cref{fig:relaxation-double-well}), where the regions corresponding to the minimum initial liquid surface area and minimal relaxation time do not overlap.

\begin{figure}[ht!]
    \centering
    \begin{subfigure}[b]{0.45\textwidth}
        \centering
        \includegraphics[width=\textwidth]{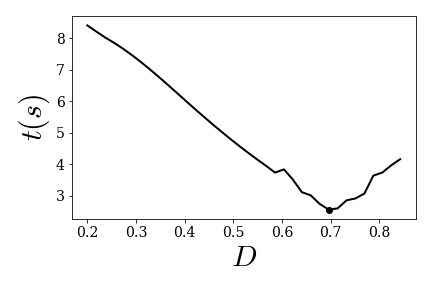}
        \caption{Relaxation time $t$ vs. double-well width $D$ ($H=0.282$).
          }
    \end{subfigure}
    \hfill
    \begin{subfigure}[b]{0.45\textwidth}
        \centering
        \includegraphics[width=\textwidth]{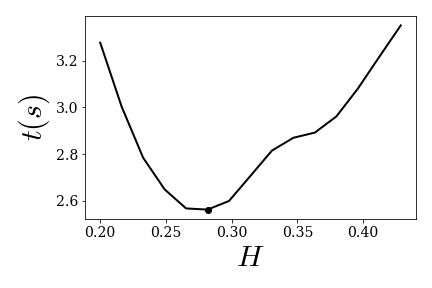}
        \caption{Relaxation time $t$ vs. indentation height $H$ ($D=0.696$).
          }
    \end{subfigure}

    \caption{One-dimensional sections through the heat map in \cref{fig:relaxation-double-well} at the optimal double-well trough parameters $H^\ast$ and $D^\ast$. Relaxation time $t$ obtained from SPH simulations as a function of (left) the double-well width $D$ ($H=0.282$) and (right) the indentation height $H$ ($D=0.696$).
    }
    
    \label{fig:relaxation-parabolic-sec}
\end{figure}

\Cref{fig:relaxation-parabolic-sec} shows one-dimensional sections through the heat map at the optimal values, $H^\ast$ and $D^\ast$. In both cases, the relaxation time exhibits a well-defined minimum, confirming that the optimum identified in the two-dimensional parameter space is locally well defined. 

\section{Conclusion}
We have investigated capillary auto-ejection as a passive method for generating relaxed centimeter-sized liquid drops during the short periods of microgravity available in drop-tower experiments. The objective was to identify cavity shapes that minimize the relaxation time of the ejected drop and thereby maximize the time available for experiments.

For parabolic troughs, analytical minimization of the initial liquid surface area predicts a unique optimal cavity shape. SPH simulations confirm that this shape also minimizes the relaxation time within the parabolic family, demonstrating a direct correspondence between surface-area minimization and dynamic relaxation.

Introducing a two-parameter family of double-well troughs provides an additional shape degree of freedom and fundamentally changes this behavior. The cavity shape yielding the shortest relaxation time no longer coincides with the one minimizing the initial liquid surface area. 

This indicates that reducing the transient deformation during ejection becomes more effective at shortening the relaxation time than further reducing the initial liquid surface area, demonstrating that optimizing the relaxation dynamics provides a more effective design criterion than surface-area minimization alone once additional shape degrees of freedom are introduced.

The observed behavior is reminiscent of the Mpemba effect \citep{Mpemba1969,Bechhoefer2021}, where a system prepared farther from equilibrium can nevertheless reach equilibrium more rapidly because it follows a more favorable dynamical pathway. Likewise, the optimal double-well trough produces a drop with a larger initial liquid surface area than the optimal parabolic trough, yet the drop relaxes significantly faster because the transient dynamics after ejection suppress shape oscillations more effectively. Although the underlying physical mechanisms are different, both phenomena illustrate that the path taken toward equilibrium may be more important than the initial distance from equilibrium.

Beyond providing practical design principles for passive drop generators intended for short-duration microgravity experiments, the present results demonstrate that introducing only a single additional shape degree of freedom fundamentally changes the optimization problem: minimizing the initial liquid surface area is no longer equivalent to minimizing the relaxation time. Extending the optimization to broader classes of cavity shapes may reveal configurations with even shorter relaxation times.



\FloatBarrier


\bibliographystyle{jfm}
\bibliography{droplet-sokolowski}

\end{document}